\documentclass{stacs_proc}

\usepackage{graphicx}
\usepackage{subfigure}
\usepackage{amsmath}
\usepackage{amsthm}
\usepackage{amsbsy}
\usepackage{amssymb}
\usepackage{amsfonts}
\usepackage[dvips]{lscape}

\DeclareMathOperator*{\argmin}{argmin}
\DeclareMathOperator*{\rank}{rank}

\DeclareMathOperator*{\card}{card}
\DeclareMathOperator*{\size}{size}
\DeclareMathOperator*{\dist}{dist}
\DeclareMathOperator*{\rad}{rad}
\DeclareMathOperator*{\diam}{diam}

\DeclareMathOperator*{\vertex}{vert}
\DeclareMathOperator*{\cost}{cost}
\DeclareMathOperator*{\vol}{vol}

\begin{document}

\title{Quantifying Homology Classes}

\author[lab1]{Chao Chen}{Chao Chen}
\author[lab1]{Daniel Freedman}{Daniel Freedman}
\address[lab1]{Rensselaer Polytechnic Institute, 110 8th Street,
Troy, NY 12180, U.S.A.}  \email[\{C. Chen, D. Freedman\}]{{chenc3,
freedman}@cs.rpi.edu}

\keywords{Computational Topology, Computational Geometry, Homology,
Persistent Homology, Localization, Optimization}
\subjclass{F.2.2, G.2.1}

\begin{abstract}
  \noindent We develop a method for measuring homology classes.  This
involves three problems.  First, we define the size of a homology
class, using ideas from relative homology. Second, we define an
optimal basis of a homology group to be the basis whose elements'
size have the minimal sum. We provide a greedy algorithm to compute
the optimal basis and measure classes in it. The algorithm runs in
$O(\beta^4 n^3 \log^2 n)$ time, where $n$ is the size of the
simplicial complex and $\beta$ is the Betti number of the homology
group. Third, we discuss different ways of localizing homology
classes and prove some hardness results.
\end{abstract}

\maketitle

\stacsheading{2008}{169-180}{Bordeaux}
\firstpageno{169}

\section{Introduction}
The problem of computing the topological features of a space has
recently drawn much attention from researchers in various
fields, such as high-dimensional data analysis
\cite{Carlsson05,Ghrist}, graphics \cite{EricksonH04,CarnerJGQ05},
networks \cite{SilvaG06} and computational biology
\cite{AgarwalEHW06,Cohen-SteinerEM06}. Topological features are often
preferable to purely geometric features, as they are more qualitative
and global, and tend to be more robust.  If the goal is to
characterize a space, therefore, features which incorporate topology
seem to be good candidates.

\begin{figure}[!btp]
     \centering
     \begin{tabular}{cccc}
     \subfigure[]{
          \label{fig:motiv1}
          \includegraphics[width=0.19\textwidth]{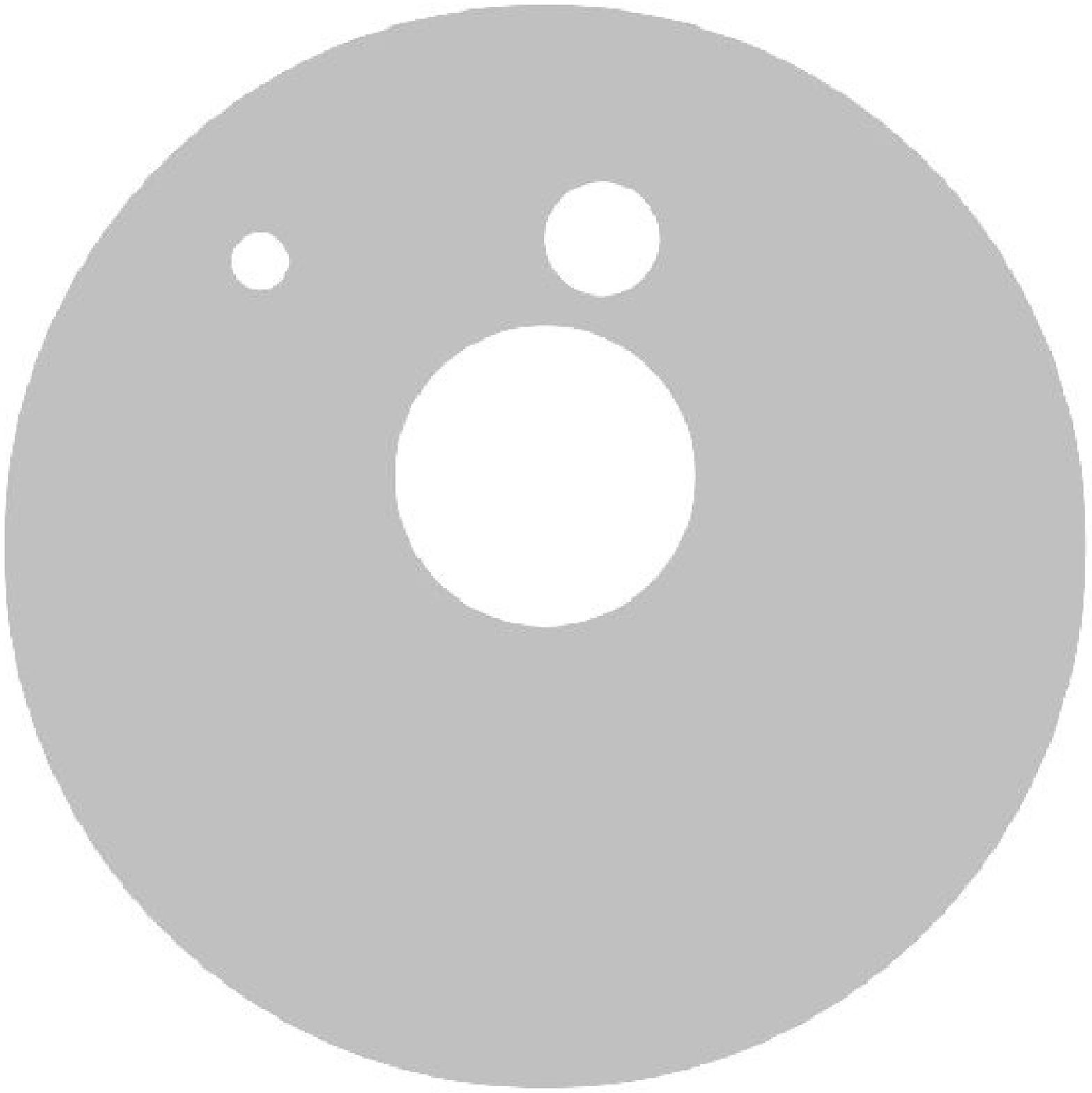}}
     &
     \subfigure[]{
          \label{fig:motiv2}
								\includegraphics[width=0.28\textwidth]{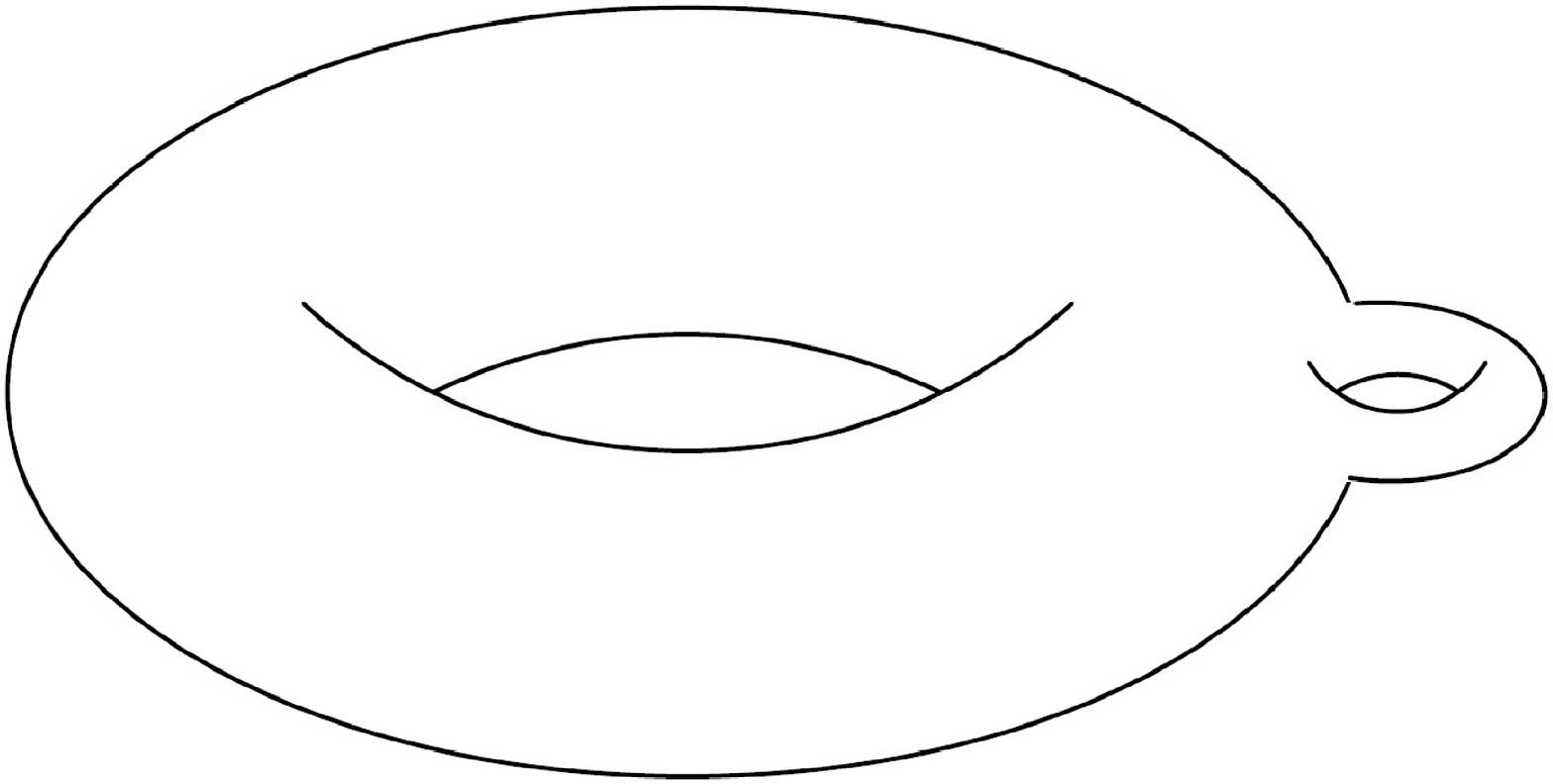}}
		 &
     \subfigure[]{
          \label{fig:motiv3}
    \includegraphics[width=0.26\textwidth]{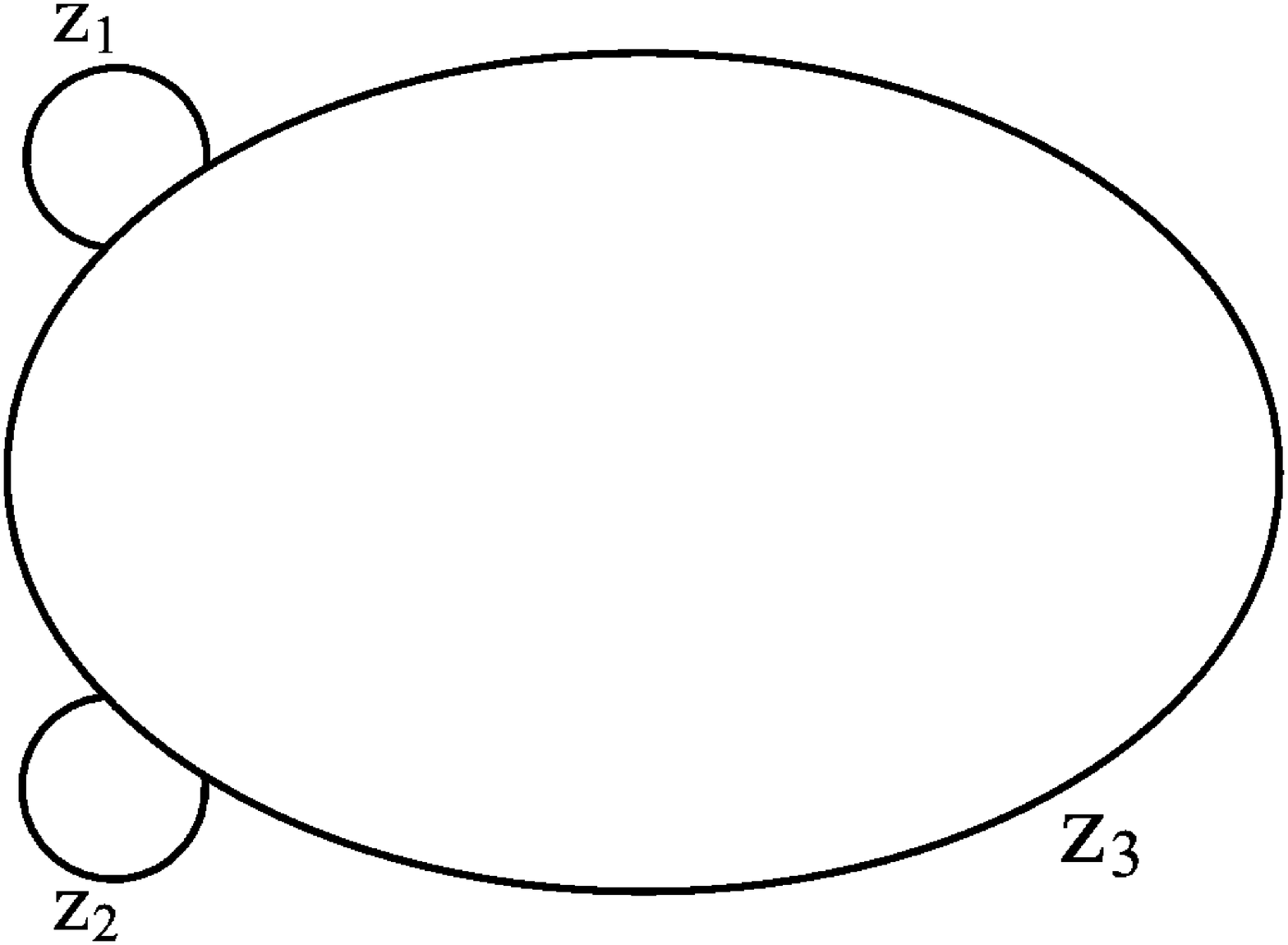}}
		 &
     \subfigure[]{
	\label{fig:motiv4}
	\includegraphics[width=0.19\textwidth]{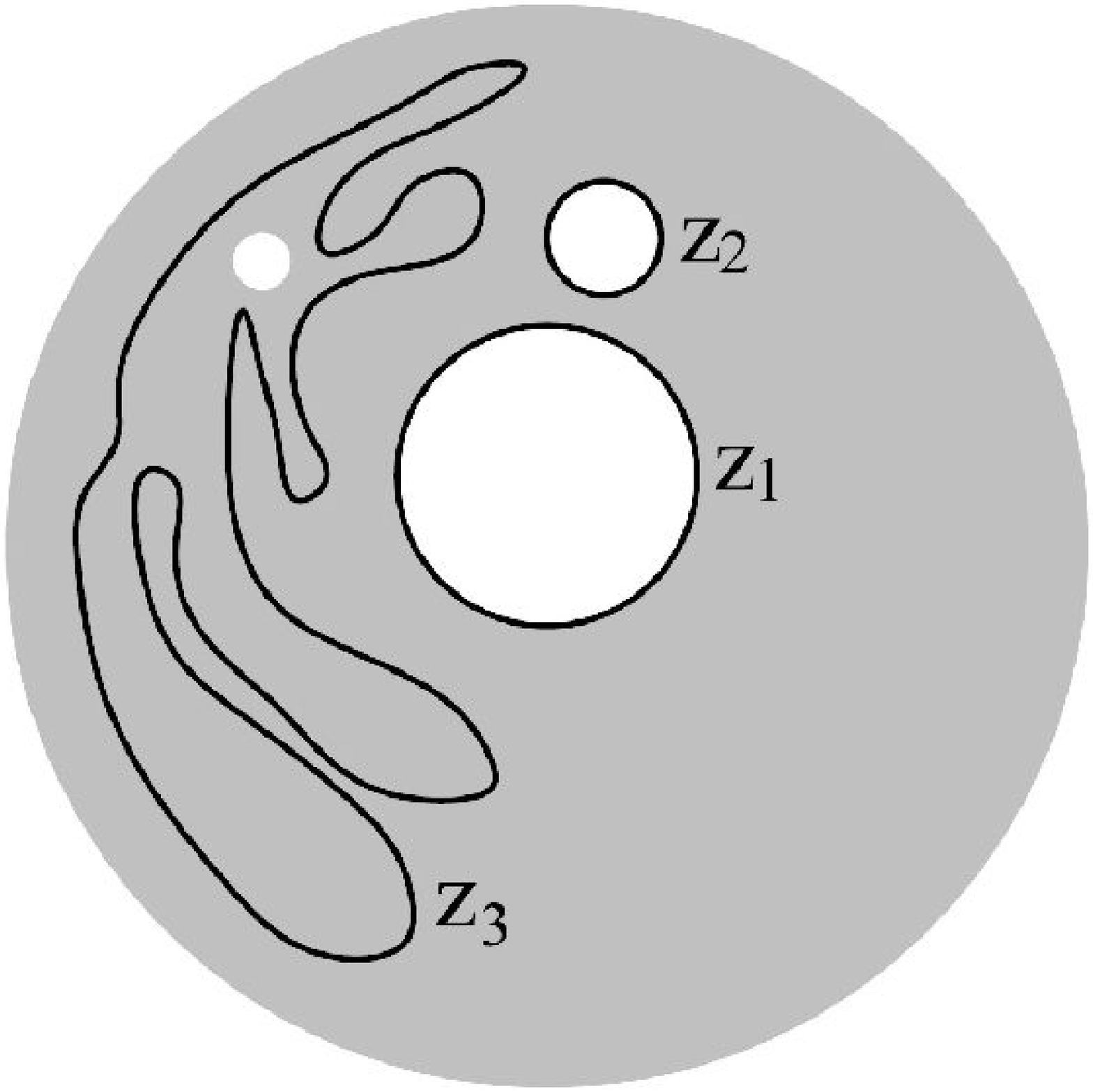}}\\
		\end{tabular}
    \caption{\small (a,b) A disk with three holes and a 2-handled
    torus are really more like an annulus and a 1-handled torus,
    respectively, because the large features are more important.  (c)
    A topological space formed from three circles.  (d) In a disk with
    three holes, cycles $z_1$ and $z_2$ are well-localized; $z_3$ is
    not.}
    \label{fig:motivation}
\end{figure}

Once we are able to compute topological features, a natural problem
is to rank the features according to their importance.  The
significance of this problem can be justified from two perspectives.
First, unavoidable errors are introduced in data acquisition, in the
form of traditional signal noise, and finite sampling of continuous
spaces.  These errors may lead to the presence of many small
topological features that are not ``real'', but are simply artifacts
of noise or of sampling \cite{WoodHDS04}.  Second, many problems are
naturally hierarchical.  This hierarchy -- which is a kind of
multiscale or multi-resolution decomposition -- implies that we want
to capture the large scale features first.  See Figure
\ref{fig:motiv1} and \ref{fig:motiv2} for examples.

The topological features we use are homology groups over
$\mathbb{Z}_2$, due to their ease of computation. (Thus, throughout
this paper, all the additions are mod 2 additions.) We would then
like to quantify or measure homology classes, as well as collections
of classes.  Specifically, there are three problems we would like to
solve:
\begin{enumerate}
\item \textbf{Measuring the size of a homology class:} We need a way
to quantify the size of a given homology class, and this size measure
should agree with intuition.  For example, in Figure
\ref{fig:motiv1}, the measure should be able to distinguish the one
large class (of the 1-dimensional homology group) from the two
smaller classes.  Furthermore, the measure should be easy to compute,
and applicable to homology groups of any dimension.

\item {\bf Choosing a basis for a homology group:}  We would like to
choose a ``good'' set of homology classes to be the generators for
the homology group (of a fixed dimension).  Suppose that $\beta$ is
the dimension of this group, and that we are using $\mathbb{Z}_2$
coefficients; then there are $2^\beta-1$ nontrivial homology classes
in total. For a basis, we need to choose a subset of $\beta$ of these
classes, subject to the constraint that these $\beta$ generate the
group.  The criterion of goodness for a basis is based on an overall
size measure for the basis, which relies in turn on the size measure
for its constituent classes.  For instance, in Figure
\ref{fig:motiv3}, we must choose three from the seven nontrivial
$1$-dimensional homology classes: $\{[z_1], [z_2], [z_3],
[z_1]+[z_2], [z_1]+[z_3], [z_2]+[z_3], [z_1]+[z_2]+[z_3]\}$.  In this
case, the intuitive choice is $\{ [z_1], [z_2], [z_3] \}$, as this
choice reflects the fact that there is really only one large cycle.

\item {\bf Localization:} We need the smallest cycle to represent a
homology class, given a natural criterion of the size of a cycle. The
criterion should be deliberately chosen so that the corresponding
smallest cycle is both mathematically natural and intuitive. Such a
cycle is a ``well-localized'' representative of its class. For
example, in Figure \ref{fig:motiv4}, the cycles $z_1$ and $z_2$ are
well-localized representatives of their respective homology classes;
whereas $z_3$ is not. 
\end{enumerate}
Furthermore, we make two additional requirements on the solution of
aforementioned problems. First, the solution ought to be computable
for topological spaces of arbitrary dimension. Second the solution
should not require that the topological space be embedded, for
example in a Euclidean space; and if the space is embedded, the
solution should not make use of the embedding. These requirements are
natural from the theoretical point of view, but may also be justified
based on real applications. In machine learning, it is often assumed
that the data lives on a manifold whose dimension is much smaller
than the dimension of the embedding space. In the study of shape, it
is common to enrich the shape with other quantities, such as
curvature, or color and other physical quantities. This leads to high
dimensional manifolds (e.g, 5-7 dimensions) embedded in high
dimensional ambient spaces \cite{CarlssonZCG05}.

Although there are existing techniques for approaching the problems
we have laid out, to our knowledge, there are no definitions and
algorithms satisfying the two requirements. 
Ordinary persistence
\cite{EdelsbrunnerLZ02,ZomorodianC05,Cohen-SteinerEH07} provides a
measure of size, but only for those \emph{inessential} classes, i.e.
classes which ultimately die. More recent work \cite{Cohen-SteinerEH}
attempts to remedy this situation, but not in an intuitive way.
Zomorodian and Carlsson \cite{ZomorodianC07} use advanced algebraic
topological machinery to solve the basis computation and localization
problems. However, both the quality of the result and the complexity
depend strongly on the choice of the given cover; there is, as yet,
no suggestion of a canonical cover. Other works like
\cite{EricksonW05,WoodHDS04,DeyLS07} are restricted to low dimension.

\paragraph{\bf Contributions.} In this paper, we solve these
problems. Our contributions include:
\begin{itemize}
\item Definitions of the size of homology classes and the optimal
homology basis.
\item A provably correct greedy algorithm to compute the optimal
homology basis and measure its classes. This algorithm uses the
persistent homology.
\item An improvement of the straightforward algorithm using finite
field linear algebra.
\item Hardness results concerning the localization of homology
classes.
\end{itemize}

\section{Defining the Problem}
In this section, we provide a technique for ranking homology
classes according to their importance. Specifically, we solve the
first two problems 
mentioned in Section 1 by formally defining (1) a meaningful size
measure for homology classes that is computable in arbitrary
dimension; and (2) an optimal homology basis which distinguishes
large classes from small ones effectively. 

Since we restrict our work to homology groups over $\mathbb{Z}_2$,
when we talk about a $d$-dimensional chain, $c$, we refer to either a
collection of $d$-simplices, or a $n_d$-dimensional vector over
$\mathbb{Z}_2$ field, whose non-zero entries corresponds to the
included $d$-simplices. $n_d$ is the number of $d$-dimensional
simplces in the given complex, $K$.
The relevant background in homology and relative homology can be
found in \cite{Munkres84}. 

\paragraph{\bf The Discrete Geodesic Distance}
In order to measure the size of homology classes, we need a notion of
distance.  As we will deal with a simplicial complex $K$, it is most
natural to introduce a discrete metric, and corresponding distance
functions.  We define the {\it discrete geodesic distance} from a
vertex $p\in \vertex(K)$, 
$f_p:\vertex(K) \to \mathbb{Z}$, as follows. For any vertex
$q\in \vertex(K)$, $f_p(q)=\dist(p,q)$ is the length of the shortest
path connecting
$p$ and $q$, in the $1$-skeleton of $K$; it is assumed that each edge
length is one, though this can easily be changed.  We may then extend
this distance function from vertices to higher dimensional simplices
naturally.  For any simplex $\sigma \in K$, $f_p(\sigma)$ is the
maximal
function value of the vertices of $\sigma$, $f_p(\sigma)=
\max_{q\in \vertex(\sigma)}f_p(q)$. 
Finally, we define a {\it discrete geodesic ball} $B_p^r$, $p\in
\vertex(K)$, $r\geq 0$, 
as the subset of $K$, $B_p^r = \{ \sigma \in K \mid f_p(\sigma) \leq
r \}$. It is straightforward to show that these subsets are in fact
{\it subcomplexes}, namely, subsets that are still simplicial
complexes.


\subsection{Measuring the Size of a Homology Class}
We start this section by introducing notions from relative homology.
Given a simplicial complex $K$ and a subcomplex $L\subseteq K$, we may
wish to study the structure of $K$ by ignoring all the chains in $L$. 
We study the {\it group of relative chain} as a quotient group,
$\mathsf{C}_d(K,L)=\mathsf{C}_d(K)/\mathsf{C}_d(L)$, whose elements
are
{\it relative chains}. Analogous to
the way we define the group of cycles $\mathsf{Z}_d(K)$, the group of
boundaries $\mathsf{B}_d(K)$ and the homology group $\mathsf{H}_d(K)$
in $\mathsf{C}_d(K)$,
we define the {\it group of relative cycles}, the {\it group of
relative boundaries} and the {\it relative homology group} in
$\mathsf{C}_d(K,L)$, denoted as $\mathsf{Z}_d(K,L)$,
$\mathsf{B}_d(K,L)$ and $\mathsf{H}_d(K,L)$, respectively.
We denote $\phi_L:\mathsf{C}_d(K)\to \mathsf{C}_d(K,L)$ as the
homomorphism mapping $d$-chains to their corresponding relative
chains, $\phi_{L}^*:\mathsf{H}_d(K)\to \mathsf{H}_d(K,L)$ as the
induced homomorphism mapping homology classes of $K$ to their
corresponding
relative homology classes.

Using these notions, we define the size
of a homology class as follows. Given a simplicial complex $ K $,
assume we are given a collection of subcomplexes 
$\mathcal{L}=\{ L \subseteq  K \}$. Furthermore, each of
these subcomplexes is endowed with a size. In this case, we define
the size of 
a homology class $h$ as the size of the smallest $ L $
carrying $h$. Here we say a subcomplex $ L $
{\it carries} $h$ if $h$ has a trivial image in the 
relative homology group $\mathsf{H}_d( K , L )$, formally,
$\phi_L^*(h)=\mathsf{B}_d( K , L )$. Intuitively, this means that $h$
disappears if we delete $L$ from K, by contracting it into a point
and modding it out. 
\begin{definition}
The size of a class $h$, $S(h)$, is the size of the smallest
measurable subcomplex carrying $h$,
formally, $S(h)=\min_{ L \in \mathcal{L}}\size( L )$ such that
$\phi_L^*(h)=\mathsf{B}_d( K , L )$.
\end{definition}

We say a subcomplex $L$ {\it carries} a chain $c$ if $L$ contains all
the simplices of the chain, formally, $c\subseteq L$. Using standard
facts from algebraic topology, it is straightforward to see that $L$
carries $h$ if and only if it carries a cycle of $h$. This gives us
more intuition behind the measure definition. 

\begin{figure}[!btp]
    \centering
    \begin{tabular}{cc}
         \subfigure[]{
          \label{fig:balls1}
		\includegraphics[width=0.2\textwidth]{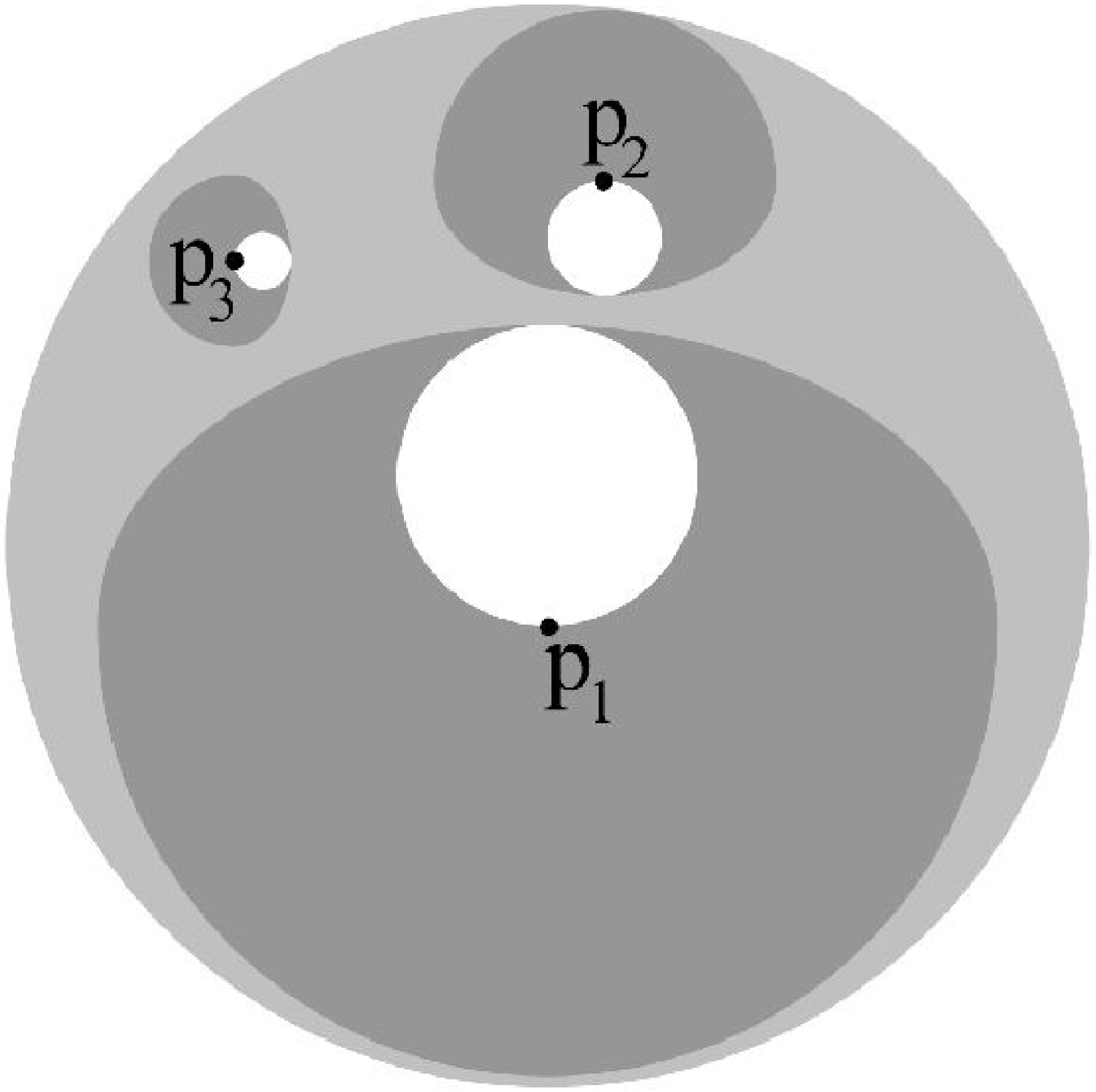}} 
		&
         \subfigure[]{
          \label{fig:balls2}
    \includegraphics[width=0.4\textwidth]{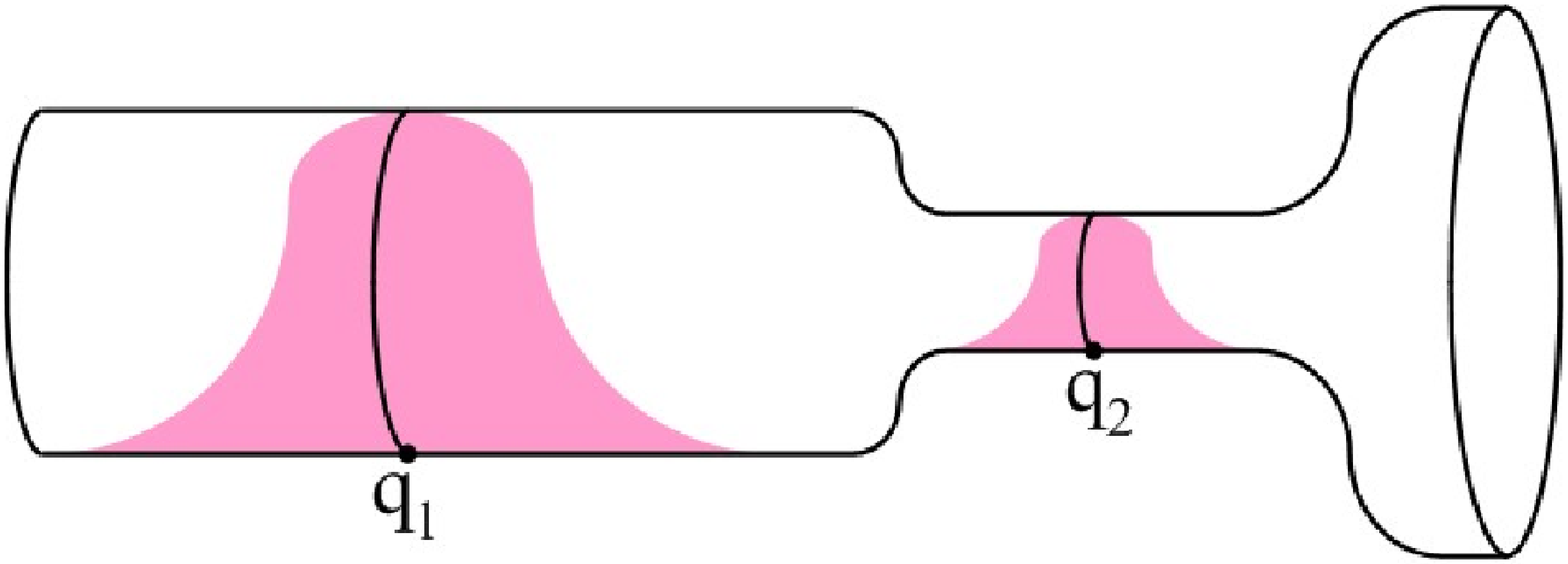}}
    \end{tabular}
    \caption{\small (a) On a disk with three holes, the three shaded
regions are the three smallest geodesic balls measuring the three
corresponding classes. (b) On a tube, the smallest geodesic ball is
centered at $q_2$, not $q_1$.}
    \label{fig:balls}
\end{figure}

In this paper, we take $\mathcal{L}$ to be the set of discrete
geodesic balls,
$\mathcal{L}=\{B_p^r\mid p\in  \vertex(K) , r\geq 0\}$.\footnote{The
idea of growing geodesic discs has been used in \cite{WoodHDS04}.
However, this work depends on low dimensional geometric reasoning,
and hence is restricted to 1-dimensional homology classes in
2-manifold.} The size of a geodesic ball is naturally its radius $r$. 
The smallest geodesic ball carrying $h$ is denoted as $B_{min}(h)$
for convenience,
whose radius is $S(h)$.
In Figure \ref{fig:balls1}, the three geodesic balls
centered at $p_1$, $p_2$ and $p_3$ are the smallest geodesic
balls carrying nontrivial homology classes $[z_1]$, $[z_2]$ and
$[z_3]$, respectively. 
Their radii are the size of the three classes. In Figure
\ref{fig:balls2}, the smallest geodesic ball 
carrying a nontrivial homology class is the pink one centered at
$q_2$,
not the one centered at $q_1$. Note that these geodesic balls may not
look like 
Euclidean balls in the embedding space.


\subsection{The Optimal Homology Basis}

For the $d$-dimensional $\mathbb{Z}_2$ homology group whose dimension
(Betti number) is $\beta_d$, there are $2^{\beta_d}-1$ nontrivial
homology classes. However, 
we only need $\beta_d$ of them to form a basis. The basis should
be chosen wisely so that we can easily distinguish important homology
classes from noise. See Figure \ref{fig:motiv3} for an example.
There are $2^3-1=7$ nontrivial homology classes; we need three of
them to form a basis. We would prefer to choose
$\{[z_1],[z_2],[z_3]\}$
as a basis, rather than $\{[z_1]+[z_2]+[z_3],[z_2]+[z_3],[z_3]\}$. 
The former indicates that there is one big cycle 
in the topological space, whereas the latter gives the impression of 
three large classes.

In keeping with this intuition, the {\it optimal homology basis}
is defined as follows.
\begin{definition}
The optimal homology basis is the basis for the homology group whose
elements' size have the minimal sum, formally,
\begin{equation*}
\mathcal{H}_d=\argmin_{\{h_1,...,h_{\beta_d}\}}\sum_{i=1}^{\beta_d}
S(h_i),s.t.
\dim(\{h_1,...,h_{\beta_d}\})=\beta_d.
\end{equation*}
\label{def:optimalBasis}
\end{definition}
This definition guarantees that large homology classes appear
as few times as possible in the optimal homology basis. 
In Figure \ref{fig:motiv3}, the optimal basis will be $\{ [z_1],
[z_2],  [z_3] \}$, which has only one large class.

For each class in the basis, we need a cycle representing it. As we
has shown, $B_{min}(h)$, the smallest geodesic ball carrying $h$,
carries at least one cycle of $h$. We localize each class in the
optimal basis by its {\it localized-cycles}, which are cycles of $h$
carried by $B_{min}(h)$. This is a fair choice because it is
consistent to the size measure of $h$ and it is computable in
polynomial time. See Section \ref{sec:loc} for further discussions.


\section{The Algorithm}
In this section, we introduce an algorithm to compute the 
optimal homology basis as defined in
Definition \ref{def:optimalBasis}. For each class in the basis, we
measure its size, and represent it with one of its localized-cycles.
We first introduce an algorithm to compute the smallest 
homology class, namely, {\sf Measure-Smallest($K$)}. Based on this
procedure, 
we provide the algorithm {\sf Measure-All($K$)}, which computes
the optimal homology basis. 
The algorithm takes $O(\beta_d^4n^4)$ time,
where $\beta_d$ is the Betti number for $d$-dimensional homology
classes and $n$ is the cardinality of the input simplicial complex
$K$. 

\paragraph{\bf Persistent Homology.} Our algorithm uses the
persistent homology algorithm. In persistent homology, we filter a
topological space with a scalar function, and capture the birth and
death times of homology classes of the sublevel set during the
filtration course. Classes with longer persistences are considered
important ones. Classes with infinite persistences are called {\it
essential homology classes} and corresponds to the intrinsic homology
classes of the given topological space. Please refer to
\cite{EdelsbrunnerLZ02,ZomorodianC05,Cohen-SteinerEH07} for theory
and algorithms of persistent homology.

\subsection{Computing the Smallest Homology Class}
\label{sec:smallestSlow}
The procedure {\sf Measure-Smallest($K$)} measures and localizes,
$h_{min}$,
the smallest nontrivial homology class, namely, the one with the
smallest
size. The output of this procedure will be a pair
$(S_{min},z_{min})$, namely,
the size and a localized-cycle
of $h_{min}$. According to the definitions, this pair is 
determined by the smallest geodesic ball carrying $h_{min}$,
namely, $B_{min}(h_{min})$. We first present the algorithm to compute
this ball. Second, we explain how to compute the pair
$(S_{min},z_{min})$
from the computed ball.

\paragraph{\bf Procedure {\sf Bmin($K$)}: Computing
$B_{min}(h_{min})$.}
It is straightforward to see that the ball $B_{min}(h_{min})$ is also
the smallest geodesic ball carrying any nontrivial homology class
of $K$. It can be computed by computing $B_p^{r(p)}$ for all vertices
$p$, where $B_p^{r(p)}$ is the smallest geodesic ball centered at $p$
which carries any nontrivial homology class. When all the
$B_p^{r(p)}$'s are computed, we compare their radii, $r(p)$'s, and
pick the smallest ball as $B_{min}(h_{min})$. 

For each vertex $p$, we compute $B_p^{r(p)}$ by applying the
persistent homology algorithm to $K$ with the discrete geodesic
distance from $p$, $f_p$, as the filter function. Note that a
geodesic ball $B_p^r$ is the sublevel set $f_p^{-1}(-\infty,
r]\subseteq K$. Nontrivial homology classes of $K$ are
essential homology classes in the persistent homology algorithm. 
(In the rest of this paper, we may use ``essential
homology classes'' and ``nontrivial homology classes of $K$''
interchangable.)
Therefore, the birth
time of the first essential homology class is $r(p)$, and the 
subcomplex $f_{p}^{-1}(-\infty,r(p)]$ is $B_{p}^{r(p)}$.

\paragraph{\bf Computing $(S_{min},z_{min})$.}
We compute the pair from the computed ball $B_{min}(h_{min})$. For
simplicity, we denote $p_{min}$ and $r_{min}$ as the center and
radius of the ball. According to the definition, $r_{min}$ is exactly
the size of $h_{min}$, $S_{min}$. 
Any nonbounding cycle (a cycle that is not a boundary) carried by
$B_{min}(h_{min})$ is a localized-cycle of $h_{min}$.\footnote{This
is true assuming that $B_{min}(h_{min})$ carries one and only one
nontrivial class, i.e. $h_{min}$ itself.  However, it is
straightforward to relax this assumption.} We first computes a basis
for all cycles carried by $B_{min}(h_{min})$, using a reduction
algorithm. Next, elements in this basis are checked one by one until
we find one which is nounbounding in $K$. This checking uses the
algorithm of 
Wiedemann\cite{Wiedemann86} for rank computation of sparse matrices
over $\mathbb{Z}_2$ field. 

\subsection{Computing the Optimal Homology Basis}

In this section, we present the algorithm for computing the optimal
homology basis defined in Definition \ref{def:optimalBasis}, namely,
$\mathcal{H}_d$. We first show that the optimal homology basis
can be computed in a greedy manner. Second, we introduce an efficient
greedy algorithm.


\subsubsection{Computing $\mathcal{H}_d$ in a Greedy Manner}
Recall that the optimal homology basis is the basis for the homology
group whose elements' size have the minimal sum.
We use matroid theory \cite{CormenLRC01} to show that 
we can compute the optimal homology basis with a greedy method.
Let $H$ be the set of nontrivial $d$-dimensional homology classes
(i.e. the homology group minus the trivial class). 
Let $L$ be the family of sets of linearly independent nontrivial
homology classes.
Then we have the following theorem, whose proof is omitted due to
space limitations.  The same 
result has been mentioned in \cite{EricksonW05}.
\begin{theorem}
The pair $(H,L)$ is a matroid when $\beta_d>0$.
\label{thm:matroid}
\end{theorem}
We construct a weighted matroid by assigning each nontrivial homology 
class its size as the weight. This weight function is strictly
positive
because a nontrivial homology class can not be carried by a geodesic
ball
with radius zero. According to matroid theory, we can compute the
optimal homology basis
with a naive greedy method: check the smallest nontrivial homology
classes one by one, until $\beta_d$ linearly independent ones are
collected. The collected $\beta_d$ classes
$\{h_{i_1},h_{i_2},...,h_{i_{\beta_d}}\}$ form the optimal
homology basis $\mathcal{H}_d$.  
(Note that the $h$'s are ordered by size, i.e. $S(h_{i_k}) \leq
S(h_{i_{k+1}})$.)
However, this method is exponential in $\beta_d$. We need a better
solution.


\subsubsection{Computing $\mathcal{H}_d$ with a Sealing Technique}

In this section, we introduce a polynomial greedy algorithm for
computing $\mathcal{H}_d$. Instead of computing the smallest classes
one by one, our
algorithm uses a sealing technique and takes time polynomial in
$\beta_d$. Intuitively, when the smallest $l$ classes in
$\mathcal{H}_d$ are picked, we make them trivial by adding new
simplices to the given complex. In the augmented complex, any linear
combinations of these picked classes becomes trivial, and the
smallest nontrivial class is the $(l+1)$'th one in $\mathcal{H}_d$.

The algorithm starts by measuring and localizing the smallest 
homology class of the given simplicial complex $K$ (using the
procedure {\sf Measure-Smallest($K$)} introduced in Section
\ref{sec:smallestSlow}), which is also the first
class we choose for $\mathcal{H}_d$. We make this class trivial by 
sealing one of its cycles -- i.e.~the localized-cycle we 
computed -- with new simplices. Next, we measure and localize the
smallest homology class of the augmented simplicial complex $K'$. 
This class is the second smallest
homology class in $\mathcal{H}_d$. We make this class trivial again
and proceed for the third smallest class in $\mathcal{H}_d$.  This
process is repeated for $\beta_d$ rounds, yielding $\mathcal{H}_d$.

\begin{figure}[!btp]
    \centering
    \begin{tabular}{cccc}
             \subfigure[]{
          \label{fig:basis1}
		\includegraphics[width=0.22\textwidth]{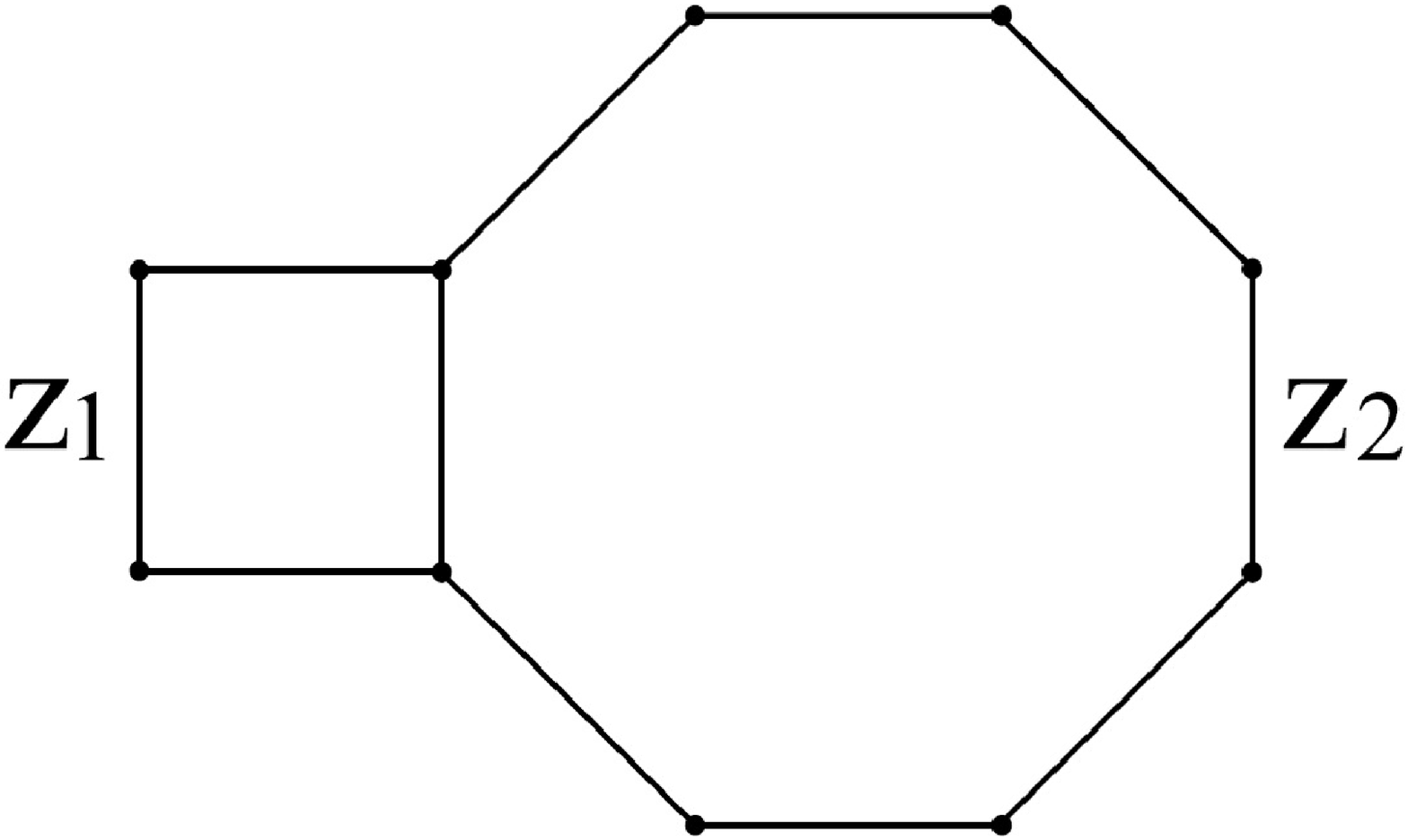}} 
		&
             \subfigure[]{
          \label{fig:basis2}
		\includegraphics[width=0.22\textwidth]{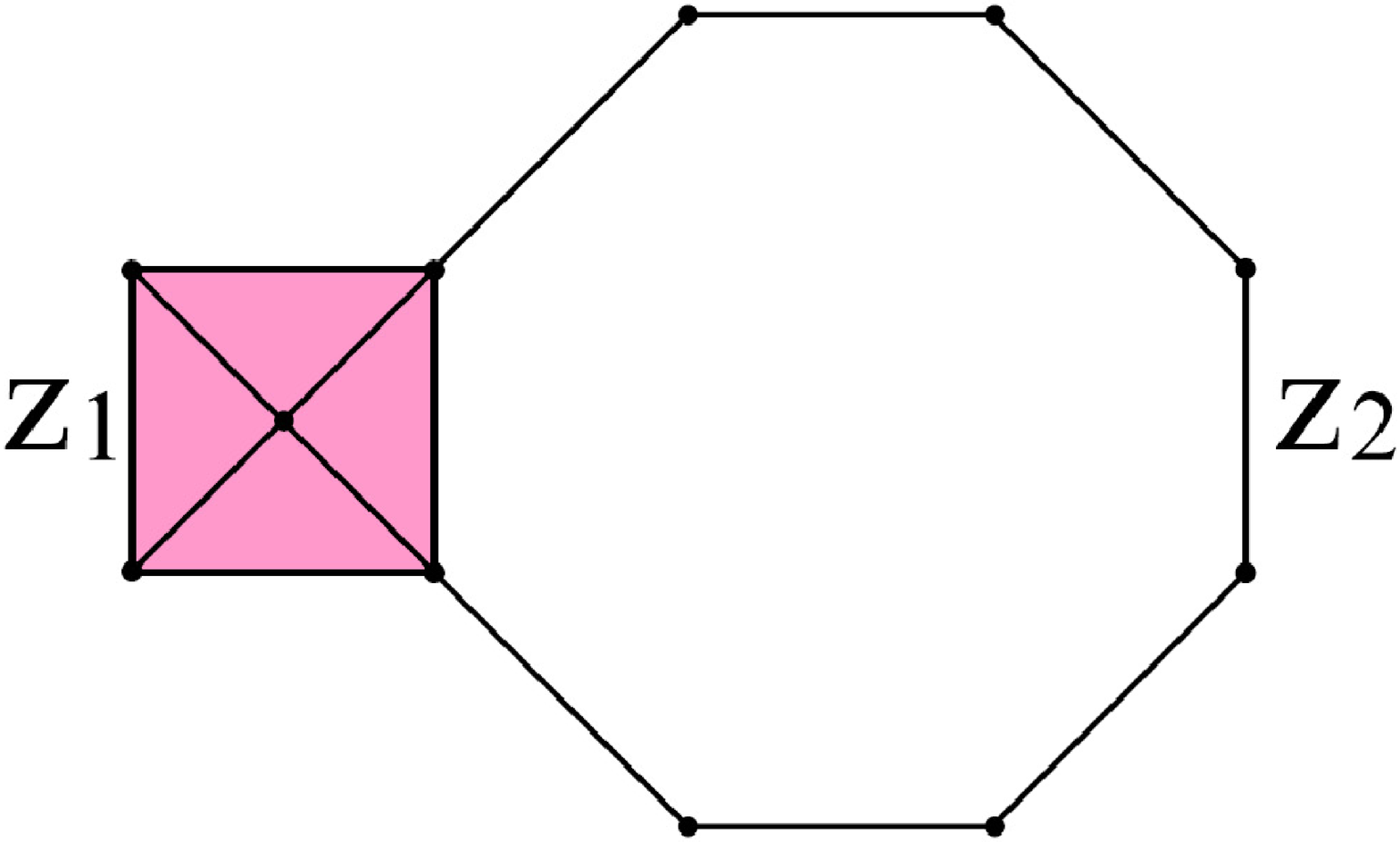}}
		& 
             \subfigure[]{
          \label{fig:seal1}
		\includegraphics[width=0.25\textwidth]{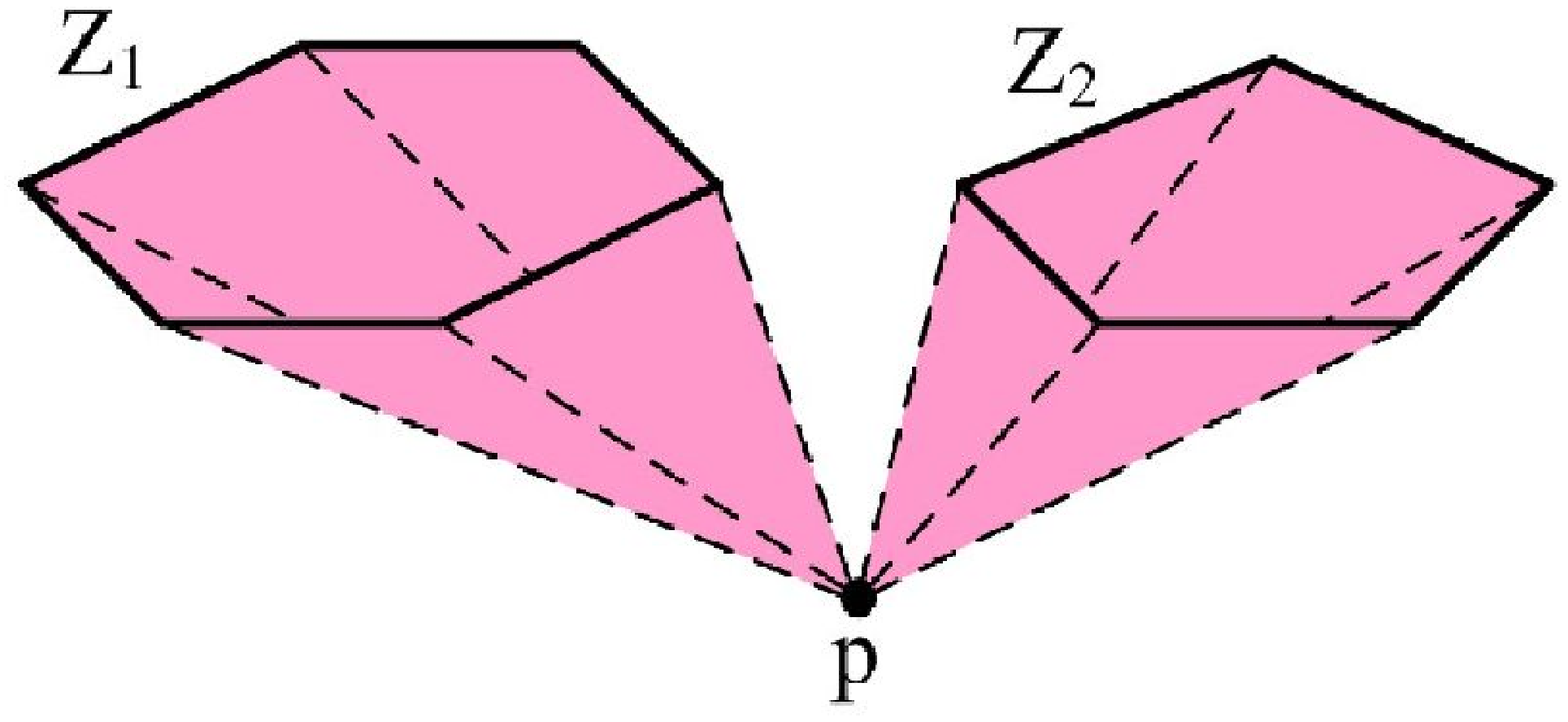}}
		& 
             \subfigure[]{
          \label{fig:seal2}
		\includegraphics[width=0.25\textwidth]{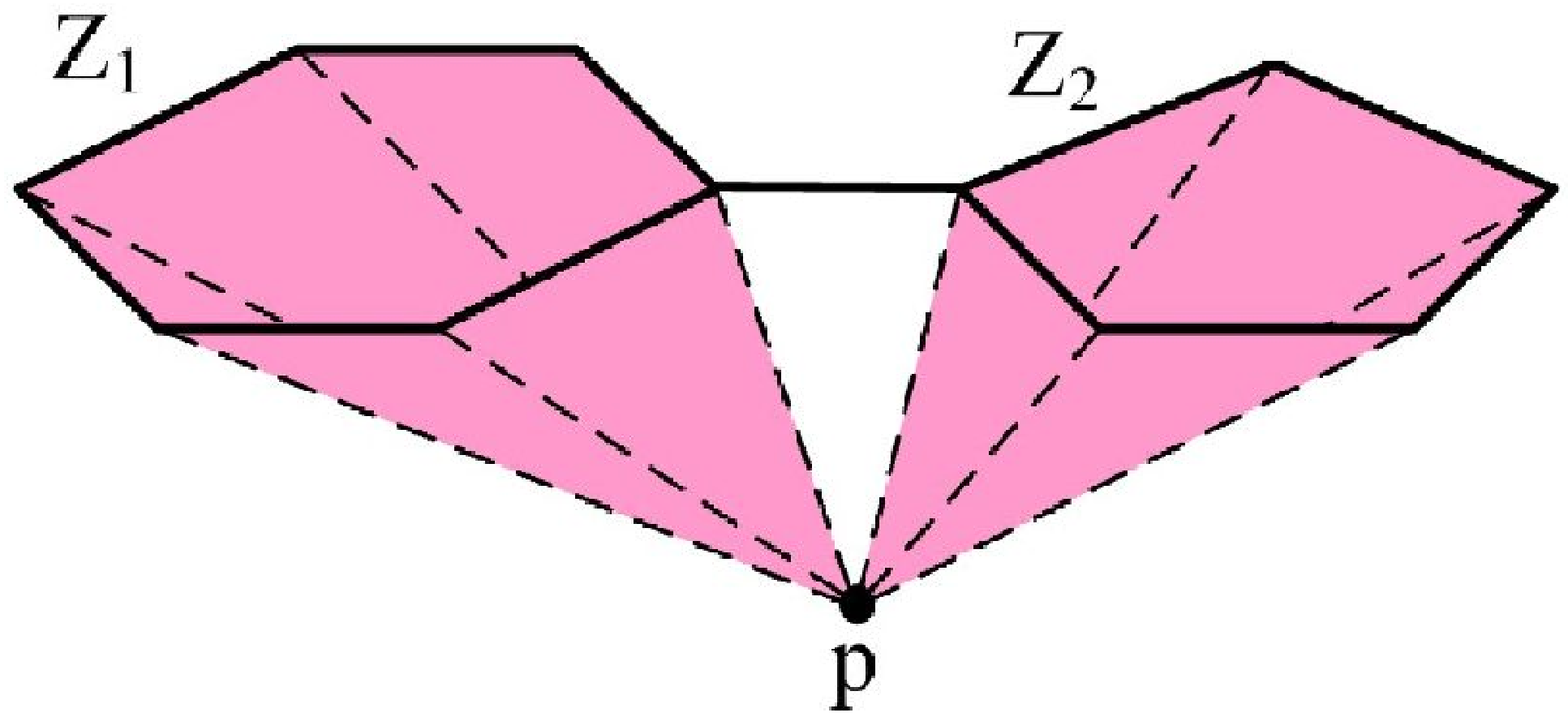}}
    \end{tabular}
    \caption{\small (a,b) the original complex $K$ and the augmented
complex $K'$ after destroying the smallest class, $[z_1]$. (c) If the
original complex $K$ consists of the two cycles $z_1$ and $z_2$,
destroying a larger class $[z_1]+[z_2]$ will make all other classes
trivial too. (d) The original complex $K$ consists of the two cycles
and an edge connecting them. Destroying $[z_1]+[z_2]$ will make all
other classes trivial and create a new class.}
    \label{fig:mAll}
\end{figure}

We make a homology class trivial by sealing the class's
localized-cycle, which we have computed. To seal this cycle $z$, we
add (a) a new vertex $v$; (b) a $(d+1)$-simplex for each $d$-simplex
of $z$, with vertex set equal to the vertex set of the $d$-simplex
together with $v$; (c) all of the faces
of these new simplices. In Figure \ref{fig:basis1} and
\ref{fig:basis2},
a $1$-cycle with four edges, $z_1$, is sealed up with one new vertex, 
four new triangles and four new edges.

It is essential to make sure the new simplices does not influence our
measurement. We assign the new vertices $+\infty$ geodesic distance
from any vertex in the original complex $K$. Furthermore, in the
procedure {\sf Measure-Smallest($K'$)}, we will not consider
any geodesic ball centered at these new vertices. In other words, the
geodesic distance from these new vertices will never be used as a
filter function.
Whenever we run
the persistent homology algorithm, all of the new simplices have
$+\infty$ 
filter function values, formally, $f_p(\sigma)=+\infty$ for all $p\in
\vertex(K)$ and $\sigma\in K'\backslash K$. 

The algorithm is illustrated in Figure \ref{fig:basis1} and
\ref{fig:basis2}. The 4-edge cycle, $z_1$, and the 8-edge cycle,
$z_2$,
are the localized-cycles of the smallest and the second smallest
homology 
classes ($S([z_1])=2$,$S([z_2])=4$).
The nonbounding cycle $z_3=z_1+z_2$ corresponds to the largest
nontrivial 
homology class $[z_3]=[z_1]+[z_2]$ ($S([z_3])=5$). After the first
round, 
we choose $[z_1]$ as the smallest class in $\mathcal{H}_1$.
Next, we destroy $[z_1]$ by sealing $z_1$, which yields the augmented 
complex $K'$. This time, we choose $[z_2]$, giving
$\mathcal{H}_1=\{[z_1],[z_2]\}$.
\paragraph{\bf Correctness.} We prove in Theorem \ref{thm:sealingUp}
the correctness of our greedy method. We begin by proving a lemma
that destroying the smallest nontrivial class will neither destroy
any other classes nor create any new classes. Please note that this
is not a trivial result. The lemma does not hold if we seal an
arbitrary class instead of the smallest one. See Figure
\ref{fig:seal1} and \ref{fig:seal2} for examples. Our proof is based
on the assumption that the smallest nontrivial class $h_{min}$ is the
only one carried by $B_{min}(h_{min})$. 
\begin{lemma}
Given a simplicial complex $K$, if we seal its smallest homology class
$h_{min}(K)$, any other nontrivial homology class of $K$, $h$, is
still
nontrivial in the augmented simplicial complex $K'$. In other words,
any cycle of $h$ is still nonbounding in $K'$.
\label{lem:sealingUp}
\end{lemma}
This lemma leads to the correctness of our algorithm, namely, Theorem
\ref{thm:sealingUp}. We prove this theorem by showing that the
procedure {\sf Measure-All($K$)} produces the same result as the
naive greedy algorithm.
\begin{theorem}
The procedure {\sf Measure-All($K$)} computes $\mathcal{H}_d$.
\label{thm:sealingUp}
\end{theorem}

\section{An Improvement Using Finite Field Linear Algebra}
In this section, we present an improvement on the algorithm
presented in the previous section, more specifically, an improvement
on computing the smallest geodesic ball carrying any nontrivial class
(the procedure {\sf Bmin}). The idea is based on
the finite field linear algebra behind the homology.

We first observe that for neighboring vertices, $p_1$ and $p_2$,
the birth times of the first essential homology class using $f_{p_1}$
and $f_{p_2}$ as filter functions are close (Theorem
\ref{thm:neighborClose}). This observation
suggests that for each $p$, instead of computing $B_{p}^{r(p)}$,
we may just test whether the geodesic ball centered at $p$ with a
certain radius carries any essential homology class.
Second, with some algebraic insight, we reduce the problem
of testing whether a geodesic ball carries any essential homology
class
to the problem of comparing dimensions of two vector spaces.
Furthermore, we use Theorem \ref{thm:cycleTest} to reduce the problem
to
rank computations of sparse matrices on the $\mathbb{Z}_2$ field,
for which we have ready tools from the literature.  In what follows,
we assume that $K$ has a single component; multiple components can be
accommodated with a simple modification.

\paragraph{\bf Complexity.} In doing so, we improve the complexity to
$O(\beta_d^4 n^3\log^2 n)$. 
More detailed complexity analysis is omitted due to space
limitations.\footnote{ This complexity is close to that of the
persistent homology algorithm, whose
complexity is $O(n^3)$. Given the nature of the problem, it seems
likely 
that the persistence complexity is a lower bound. If this is the
case, the current
algorithm is nearly optimal. Cohen-Steiner et
al.\cite{Cohen-SteinerEM06}
provided a linear algorithm to maintain the persistences while
changing
the filter function. While interesting, this algorithm is not
applicable in our case. }

Next, we present details of the improvement. In Section
\ref{sec:improve1}, we prove Theorem \ref{thm:neighborClose} and
provide details of the improved algorithm. In Section
\ref{sec:improve2}, we explain how to test whether a certain
subcomplex carries any essential homology class of $K$. For
convenience, in this section, we use ``carrying nonbounding cycles''
and ``carrying essential homology classes'' interchangeably, because
a geodesic ball
carries essential homology classes of $K$ if and only if it carries
nonbounding
cycles of $K$.


\subsection{The Stability of Persistence Leads to An Improvement}
\label{sec:improve1}
Cohen-Steiner et al.\cite{Cohen-SteinerEH07} proved that the change,
suitably defined, of the persistence of homology classes is bounded
by the changes of the filter functions. Since the filter functions of
two neighboring vertices, $f_{p_1}$ and $f_{p_2}$,
are close to each other, the birth times of the first nonbounding
cycles in
both filters are close as well. This leads to Theorem
\ref{thm:neighborClose}. A simple proof is provided.
\begin{theorem}
If two vertices $p_1$ and $p_2$ are neighbors, the birth times of the
first
nonbounding cycles for filter functions $f_{p_1}$ and 
$f_{p_2}$ differ by no more than 1.
\label{thm:neighborClose}
\end{theorem}
\proof
$p_1$ and $p_2$ are neighbors implies that for any point $q$, 
$f_{p_2}(q)\leq f_{p_2}(p_1)+f_{p_1}(q)=1 + f_{p_1}(q)$,
in which the inequality follows the triangular inequality. Therefore,
$B_{p_1}^{r(p_1)}$ is a subset of $B_{p_2}^{r(p_1)+1}$. The former
carries nonbounding cycles implies that the latter does too, and thus
$r(p_2)\leq r(p_1)+1$. Similarly, we have $r(p_1)\leq r(p_2)+1$.
\qed

This theorem suggests a way to avoid computing $B_{p}^{r(p)}$
for all $p\in \vertex(K)$ in the procedure {\sf Bmin}.
Since our objective is to find the minimum of the $r(p)$'s, we do a 
breadth-first search through all the vertices with 
global variables $r_{min}$ and $p_{min}$ recording the smallest
$r(p)$ we have found and its corresponding center $p$, respectively. 
We start by applying the persistent homology algorithm on $K$ with
filter function $f_{p_0}$, where $p_0$ is an arbitrary vertex of $K$.
Initialize $r_{min}$ as the birth time
of the first nonbounding cycle of $K$, $r(p_0)$, and $p_{min}$ as
$p_0$. 
Next, we do a breadth-first
search through the rest vertices. For each vertex
$p_i, i\neq 0$, there is a neighbor $p_j$ we have visited (the parent
vertex of $p_i$ in the breath-first search tree). We know that
$r(p_j)\geq r_{min}$ and $r(p_i)\geq r(p_j)-1$ (Theorem
\ref{thm:neighborClose}).
Therefore, $r(p_i)\geq r_{min}-1$.
We only need to test whether the geodesic ball
$B_{p_i}^{r_{min}-1}$ carries any nonbounding cycle of $K$. If so,
$r_{min}$ is decremented by one, and $p_{min}$ is updated to $p_i$.
After all vertices are visited, $p_{min}$ and $r_{min}$ give us the
ball we want.

\begin{figure}[!btp]
    \centering
    \begin{tabular}{cccc}
    \subfigure[]{
    \label{fig:tTail1}
		\includegraphics[width=0.25\textwidth]{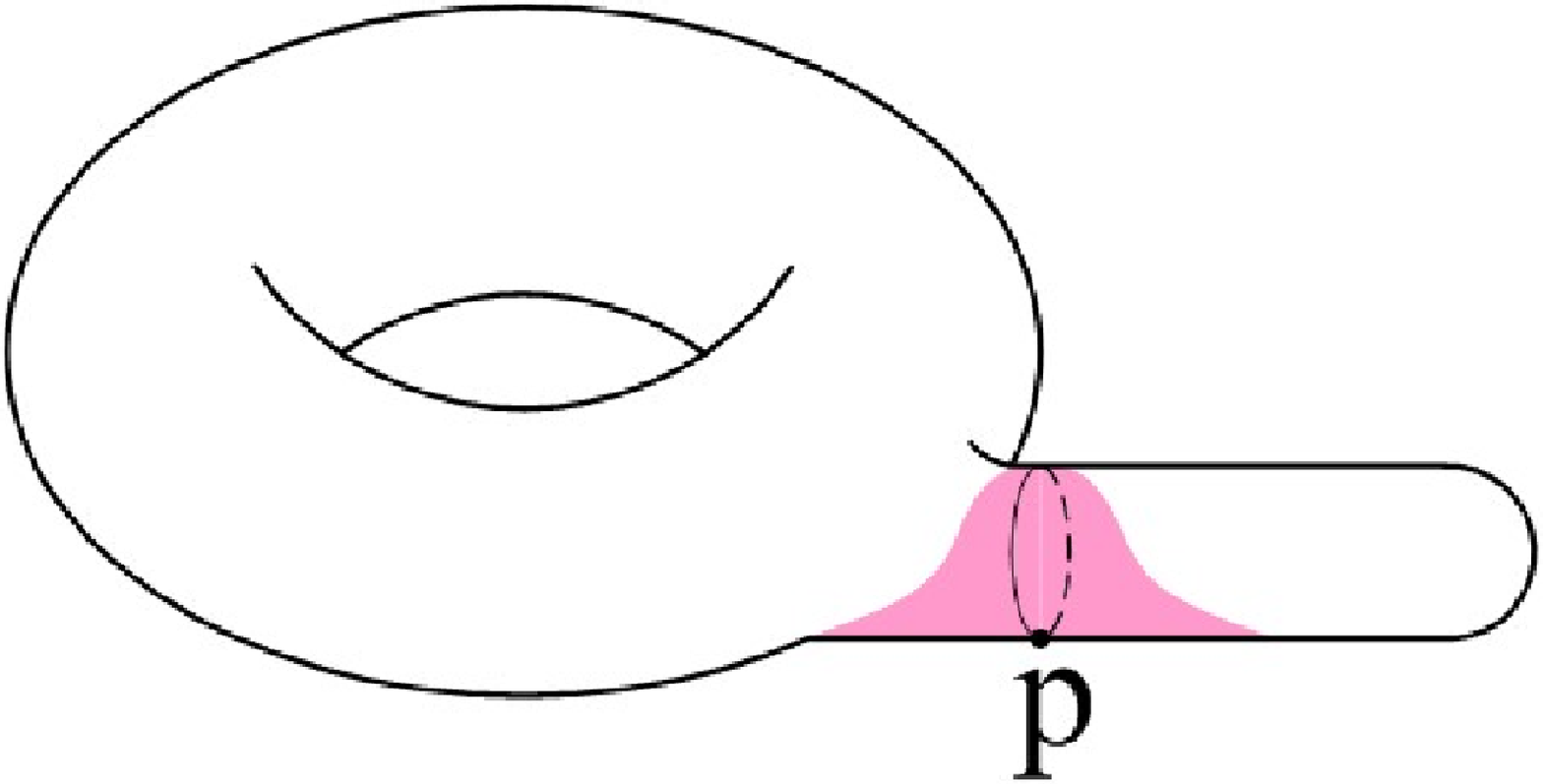}} 
		& 
		\subfigure[]{
		\label{fig:tTail2}
		\includegraphics[width=0.25\textwidth]{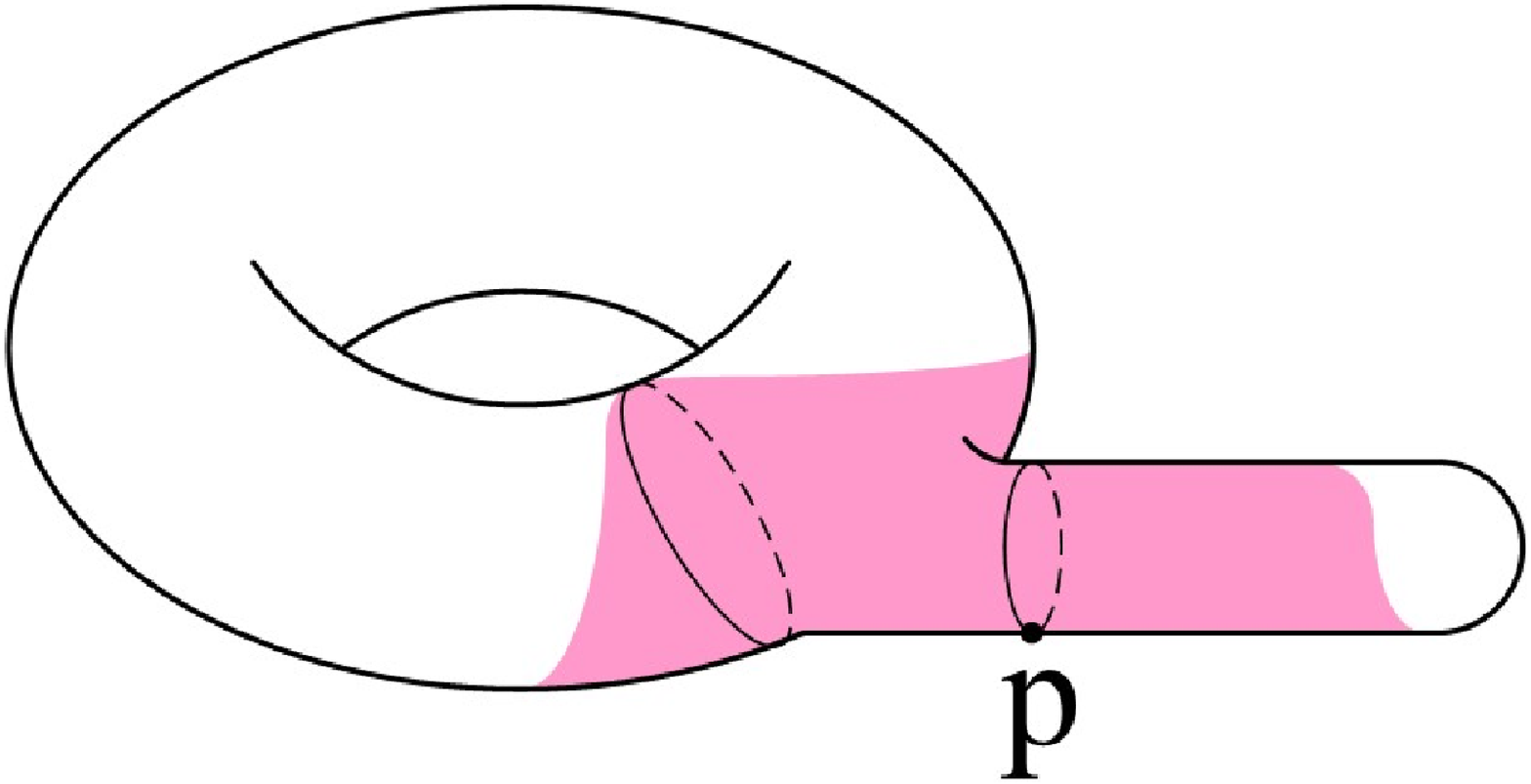}}
    &
		\subfigure[]{
		\label{fig:wiggle1}
    \includegraphics[width=0.2\textwidth]{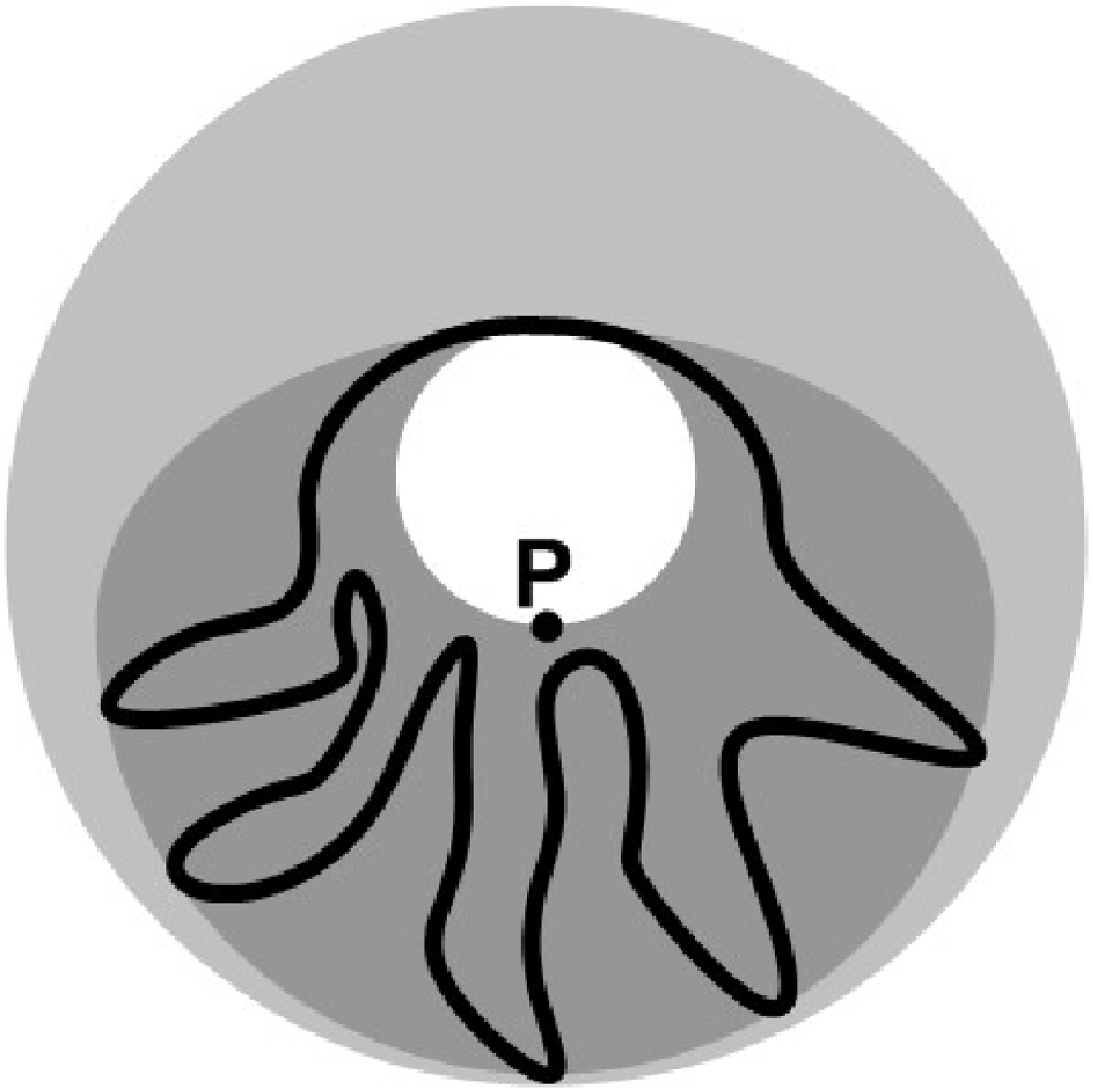}}
		&    
		\subfigure[]{
		\label{fig:wiggle2}
    \includegraphics[width=0.2\textwidth]{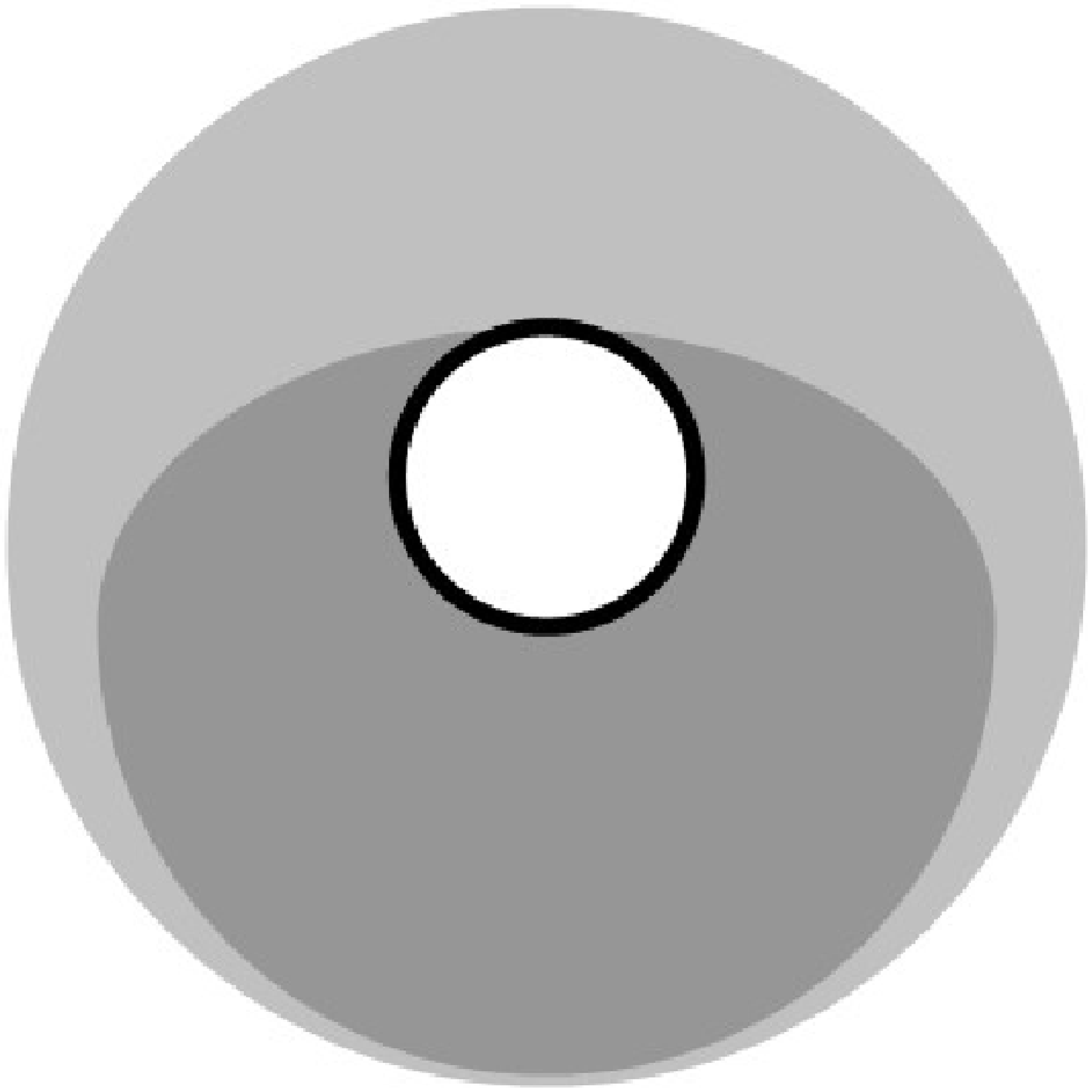}}
    \end{tabular}
    \caption{\small (a,b) In a torus with a tail, only the ball in the
    second figure carries nonbounding cycles of $K$, although in both
figures the balls have nontrivial topology. (c,d) The cycles with the
minimal radius and the minimal diameter, $z_r$ and $z_d$ (Used in
Section \ref{sec:loc}).}
    \label{fig:tTail}
\end{figure}

However, testing whether the subcomplex $B_{p_i}^{r_{min}-1}$ carries
any
nonbounding cycle of $K$ is not as easy as computing nonbounding
cycles 
of the subcomplex. A nonbounding cycle of $B_{p}^{r_{min}-1}$
may not be nonbounding in $K$ as we require.
For example, in Figure \ref{fig:tTail1} and \ref{fig:tTail2}, the
simplicial complex $K$ is a torus with a tail. The pink geodesic ball
in the first figure does not carry any nonbounding cycle of $K$,
although it carries its own nonbounding cycles. The geodesic ball in
the second figure is the one that carries nonbounding cycles of $K$.
Therefore, we need algebraic tools to distinguish nonbounding cycles
of $K$ from those of the subcomplex $B_{p_i}^{r_{min}-1}$.

\subsection{Procedure {\sf Contain-Nonbounding-Cycle}: Testing
Whether a \label{sec:improve2}
Subcomplex Carries Nonbounding Cycles of $K$}
In this section, we present the procedure for testing whether 
a subcomplex $K_0$ carries any nonbounding cycle of $K$. 
A chain in $K_0$ is a cycle if and only if it is a cycle of $K$.
However, solely from $K_0$, we are not able to tell whether a
cycle carried by $K_0$ bounds or not in $K$.
Instead, we write the set of cycles of $K$ carried by $K_0$, 
$\mathsf{Z}_d^{K_0}(K)$, and the set
of boundaries of $K$ carried by $K_0$, $\mathsf{B}_d^{K_0}(K)$,
as sets of linear combinations with certain constraints.
Consequently, 
we are able to test 
whether any cycle carried by $K_0$ is nonbounding in $K$ by comparing
their dimensions. Formally, we define
$\mathsf{B}_d^{K_0}(K)=\mathsf{B}_d(K)\cap \mathsf{C}_d(K_0)$ and
$\mathsf{Z}_d^{K_0}(K)=\mathsf{Z}_d(K)\cap \mathsf{C}_d(K_0)$. 

Let $\hat{H}_d=[z_1,...,z_{\beta_d}]$ be the matrix whose column
vectors are arbitrary $\beta_d$ nonbounding cycles of $K$ which are
not homologous to each other. The boundary group and the cycle group
of $K$ are column spaces of the matrices $\partial_{d+1}$ and
$\hat{Z}_d=[\partial_{d+1},\hat{H}_d]$, respectively.
Using finite field linear algebra, we have the following theorem,
whose proof is omitted due to space limitations. 
\begin{theorem}
\label{thm:cycleTest}
$K_0$ carries nonbounding cycles of $K$ if and only if 
\begin{eqnarray*}
\rank(\hat{Z}_d^{K\backslash K_0})-\rank(\partial_{d+1}^{K\backslash
K_0}) \neq \beta_d.
\end{eqnarray*}
where $\partial_{d+1}^i$ and $\hat{Z}_d^i$ are the $i$-th rows of the
matrices 
$\partial_{d+1}$ and $\hat{Z}_d$, respectively.
\end{theorem}
We use the algorithm of Wiedemann\cite{Wiedemann86} for the rank
computation. 
In our algorithm, the boundary matrix $\partial_{d+1}$ is 
given. The matrix $\hat{H}_d$ can be precomputed as follows.
We perform a column reduction on the boundary matrix
$\partial_d$ to compute a basis for the cycle group
$\mathsf{Z}_d(K)$. 
We check elements in this basis one by one until we collect $\beta_d$
of them forming $\hat{H}_d$. For each cycle $z$ in this cycle basis,
we check whether $z$ is linearly independent of the $d$-boundaries
and the nonbounding cycles we have already chosen. More details are
omitted due to space limitations.

\section{Localizing Classes}
\label{sec:loc}
In this section, we address the localization problem. We formalize
the localization problem as a combinatorial optimization problem:
Given a simplcial complex $K$, compute the representative cycle of a
given homology class minimizing a certain objective function.
Formally, given an objective function defined on all the cycles,
$\cost:\mathsf{Z}_d(K)\to \mathbb{R}$, we want to localize a given
class with its {\it optimally localized cycle},
$z_{opt}(h)=\argmin_{z\in h}\cost(z)$. In general, we assume the
class $h$ is given by one of its representative cycles, $z_0$.

We explore three options of the objective function $\cost(z)$,
i.e.~the {\it volume}, {\it diameter} and {\it radius} of a given
cycle $z$. We show that the cycle with the minimal volume and the
cycle with the minimal diameter are NP-hard to compute. The cycle
with the minimal radius, which is the localized-cycle we defined and
computed in previous sections, is a fair choice. Due to space
limitations, we omit proofs of theorems in this section.
\begin{definition}[Volume]
\label{def:vol}
The volume of $z$ is the number of its simplices, $\vol(z)=\card(z)$.
\end{definition}
For example, the volume of a 1-dimensional cycle, a 2-dimensional
cycle and a 3-dimensional cycle are the numbers of their edges,
triangles and tetrahedra, respectively. A cycle with the smallest
volume, denoted as $z_v$, is consistent to a ``well-localized'' cycle
in intuition. Its 1-dimensional version, the shortest cycle of a
class, has been studied by researchers
\cite{EricksonW05,WoodHDS04,DeyLS07}. However, we prove in Theorem
\ref{thm:optVol} that computing $z_v$ of $h$ is
NP-hard.\footnote{Erickson and Whittlesey \cite{EricksonW05}
localized 1-dimensional classes with their shortest representative
cycles. Their polynomial algorithm can only localize classes in the
shortest homology basis, not arbitrary given classes.} The proof is
by reduction from the NP-hard problem MAX-2SAT-B
\cite{PapadimitriouY88}. More generally, we can extend the the volume
to be the sum of the weights assigned to simplices of the cycle,
given an arbitrary weight function defined on all the simplices of
$K$. The corresponding smallest cycle is still NP-hard to compute.
\begin{theorem}
\label{thm:optVol}
Computing $z_v$ for a given $h$ is NP-hard.
\end{theorem}

When it is NP-hard to compute $z_v$, one may resort to the geodesic
distance between elements of $z$. The second choice of the objective
function is the diameter.
\begin{definition}[Diameter]
\label{def:diam}
The diameter of a cycle is the diameter of its vertex set,
$\diam(z)=\diam(\vertex(z))$, in which the diameter of a set of
vertices is the maximal geodesic distance between them, formally,
$\diam(S)=\max_{p,q\in S}\dist(p,q)$.
\end{definition}
Intuitively, a representative cycle of $h$ with the minimal diameter,
denoted $z_d$, is the cycle whose vertices are as close to each other
as possible. The intuition will be further illustrated by comparison
against the radius criterion. We prove in Theorem \ref{thm:optDiam}
that computing $z_d$ of $h$ is NP-hard, by reduction from the NP-hard
{\it Multiple-Choice Cover Problem} (MCCP) of Arkin and Hassin
\cite{ArkinH00}.
\begin{theorem}
\label{thm:optDiam}
Computing $z_d$ for a given $h$ is NP-hard.
\end{theorem}

The third option of the objective function is the radius.
\begin{definition}[Radius]
The radius of a cycle is the radius of the smallest geodesic ball
carrying it, formally, $\rad(z)=\min_{p\in \vertex(K)}\max_{q\in
\vertex(z)}\dist(p,q)$, where $\vertex(K)$ and $\vertex(z)$ are the
sets of vertices of the given simplicial complex $K$ and the cycle
$z$, respectively.
\end{definition}
The representative cycle with the minimal radius, denoted as $z_r$,
is the same as the localized-cycle defined and computed in previous
sections. Intuitively, $z_r$ is the cycle whose vertices are as close
to a vertex of $K$ as possible. However, $z_r$ may not necessarily be
localized in intuition. It may wiggle a lot while still being carried
by the smallest geodesic ball carrying the class. See Figure
\ref{fig:wiggle1}, in which we localize the only nontrivial homology
class of an annulus (the light gray area). The dark gray area is the
smallest geodesic ball carrying the class, whose center is $p$.
Besides, the cycle with the minimal diameter (Figure
\ref{fig:wiggle2}) avoids this wiggling problem and is concise in
intuition. This in turn justifies the choice of
diameter.\footnote{This figure also illustrates that the radius and
the diameter of a cycle are not strictly related. For the cycle $z_r$
in the left, its diameter is twice of its radius. For the cycle $z_d$
in the center, its diameter is equal to its radius.} 
We can prove that $z_r$ can be computed in polynomial time and is a
2-approximation of $z_d$. 
\begin{theorem}
We can compute $z_r$ in polynomial time.
\end{theorem}
\begin{theorem}
\label{thm:radApxDiam}
$\diam(z_r) \leq 2\diam(z_d)$.
\end{theorem}
This bound is a tight bound. In Figure \ref{fig:wiggle1} and
\ref{fig:wiggle2}, the diameter of the cycle $z_r$ is twice of the
radius of the dark gray geodesic ball. The diameter of the cycle
$z_d$ is the same as the radius of the ball. We have
$\diam(z_r)=2\diam(z_d)$.

\vskip-0.3cm
\section*{Acknowledgements}
  The authors wish to acknowledge constructive comments from
anonymous reviewers and fruitful discussions on persistent homology
with Professor Herbert Edelsbrunner. 

\bibliographystyle{abbrv}

\end{document}